\documentclass[nohyper,12pt,a4paper]{JHEP}
\usepackage{epsfig}
\usepackage{psfrag}
\usepackage{graphicx}
\usepackage{amsfonts,amsmath,amssymb}

\title{Holographic Superconductors with Lifshitz Scaling}

\author{E. J. Brynjolfsson,$^{1)}$ U. H. Danielsson,$^{2)}$, L. Thorlacius,$^{1),3)}$
and T. Zingg$^{1),3)}$\\
1) University of Iceland, Science Institute \\
Dunhaga 3, IS-107 Reykjavik, Iceland\\
\ \\
2) Institutionen f\"or fysik och astronomi\\
Uppsala Universitet, Box 803, SE-751 08 Uppsala, Sweden\\
\ \\
3) NORDITA, Roslagstullsbacken 23 \\
SE-106 91 Stockholm, Sweden \\ 
\ \\
E-mail: \email{erlingbr@hi.is}, \email{ulf.danielsson@physics.uu.se}, 
\email{lth@nordita.org}, \email{zingg@nordita.org}}

\abstract{Black holes in asymptotically Lifshitz spacetime provide a window onto 
finite temperature effects in strongly coupled Lifshitz models. We add a Maxwell gauge 
field and charged matter to a recently proposed gravity dual of 2+1 dimensional Lifshitz 
theory. This gives rise to charged black holes with scalar hair, which correspond to the 
superconducting phase of holographic superconductors with $z > 1$ Lifshitz scaling.  
Along the way we analyze the global geometry of static, asymptotically Lifshitz black holes 
at arbitrary critical exponent $z>1$. In all known exact solutions there is a null curvature 
singularity in the black hole region, and, by a general argument, the same applies 
to generic Lifshitz black holes.}


\preprint{NORDITA-2009-46, UUITP-19/09}

\begin{document}


\section{Introduction}

A holographic superconductor \cite{Gubser:2008px,Hartnoll:2008vx} is dual to an 
asymptotically AdS spacetime with a charged black hole carrying scalar hair. 
It is a theoretical construction where a superconducting phase transition in a strongly 
coupled system is studied via the AdS/CFT correspondence \cite{Maldacena:1997re}. 
Having a black hole places the system at finite temperature, the black hole charge 
gives rise to a chemical potential in the dual theory, and the scalar field hair signals
the condensation of a charged operator in the dual theory.  
Without a chemical potential all temperatures are equivalent due to the 
underlying conformal symmetry and there can be no phase transition. 
With a chemical potential, on the other hand, a new scale is introduced that 
allows for a phase transition at some critical temperature. 

The black holes in question are found as solutions of AdS gravity coupled to
a Maxwell gauge field and matter in the form of a charged scalar. Black holes that 
are neutral under the Maxwell field do not develop any hair. Near a charged
black hole, however, the gauge field sourced by the black hole couples to the
charged scalar and induces a negative mass squared sufficient for condensation
provided the temperature is low enough. It was observed in \cite{Hartnoll:2008kx} that 
even a neutral scalar can lead to condensation below a non zero critical temperature 
due to the existence of a new and effectively lower Breitenlohner-Freedman 
bound \cite{Breitenlohner:1982jf} near the horizon of a low temperature, near extremal, 
AdS-Reissner-Nordstr\"{o}m black hole. See \cite{Hartnoll:2009sz} and 
\cite{Herzog:2009xv} for recent reviews of holographic superconductivity.

In the present paper we show that similar phenomena can occur in a theory
exhibiting Lifshitz scaling,
\begin{equation}
t\rightarrow\lambda^{z}t,\qquad\mathbf{x}\rightarrow\lambda\mathbf{x},
\label{lscaling}%
\end{equation}
with $z\neq1$. Models with scaling of this type have, for example, been used to 
model quantum critical behavior in strongly correlated electron systems
\cite{Rokhsar-1988}-\cite{vishwanath-2003}.
Our starting point is the model proposed by Kachru, Liu,
and Mulligan \cite{Kachru:2008yh} for the holographic study of strongly coupled 
2+1 dimensional systems with Lifshitz scaling. While this model in itself does not 
support a phase transition to a superconductor it turns out that a relatively simple 
extension does.

A quantum critical point exhibiting dynamical scaling of the form (\ref{lscaling}) 
has a gravitational dual description in terms of a spacetime metric of the form
\begin{equation}
ds^{2}=L^{2}\left(  -r^{2z}dt^{2}+\frac{dr^{2}}{r^{2}}+r^{2}d^{2}%
\mathbf{x}\right)  ,\label{lmetric}%
\end{equation}
which is invariant under the transformation
\begin{equation}
t\rightarrow\lambda^{z}t,\quad r\rightarrow\frac{r}{\lambda},\quad
\mathbf{x}\rightarrow\lambda\mathbf{x}\,.
\end{equation}
Here $L$ is a characteristic length scale and the coordinates 
$(t,r,x^{1},x^{2})$ are taken to be dimensionless. Critical points with $z>1$ are
often said to be non-relativistic. This can be motivated by considering null
geodesics at fixed $r=r_0$ in the Lifshitz background (\ref{lmetric}). Along 
such a geodesic one finds that
$(d\mathbf{x}/dt)^2=r_0^{2z-2}$, and so for $z>1$ the effective speed of light in the 
boundary theory diverges as $r_0\rightarrow\infty$.

In \cite{Kachru:2008yh} it
was shown how the fixed point geometry (\ref{lmetric}), with $z>1$, can be 
obtained from an action coupling 3+1 dimensional gravity with a negative 
cosmological constant to abelian two- and three-form field strengths,
\begin{align}
S  &  =\int\mathrm{d}^{4}x\sqrt{-g}\;\left(  R-2\Lambda\right)
\label{actionform}\\
&  -\frac{1}{2}\int\ast F_{(2)}\wedge F_{(2)}-\frac{1}{2}\int\ast
H_{(3)}\wedge H_{(3)}-c\int B_{(2)}\wedge F_{(2)}\nonumber
\end{align}
with $H_{(3)}=d B_{(2)}$ and the cosmological constant and the coupling 
between the form-fields given
by $\Lambda=-\frac{z^{2}+z+4}{2L^{2}}$ and $c=\frac{\sqrt{2z}}{L}$.
The equations of motion for the form fields can be written
\begin{equation}
\label{formeq}
d\ast F_{\left(  2\right)  }  =-cH_{\left(  3\right)  }, \qquad
d\ast H_{\left(  3\right)  }  =-cF_{\left(  2\right)  },
\end{equation}
and the Einstein equations are
\begin{equation}
\label{einsteineqs}
G_{\mu\nu}+\Lambda g_{\mu\nu}=\frac{1}{2}(F_{\mu\lambda}F_{\nu
}^{\>\lambda}-\frac{1}{4}g_{\mu\nu}F_{\lambda\sigma}F^{\lambda\sigma}
)+\frac{1}{4}(H_{\mu\lambda\sigma}H_{\nu}^{\>\lambda\sigma}-\frac{1}{6}
g_{\mu\nu}H_{\lambda\sigma\rho}H^{\>\lambda\sigma\rho}).
\end{equation}

Black holes in this 3+1 dimensional gravity theory were considered in
\cite{Danielsson:2009gi}, where numerical black hole solutions were found at 
$z=2$ and used to study finite temperature effects in the dual 2+1 dimensional
system. Related work on Lifshitz black holes can be found in \cite{Mann:2009yx}, 
where topological black holes with hyperbolic horizons were included, and in
\cite{Bertoldi:2009vn,Bertoldi:2009dt}, where black holes at general values of $z$ 
were considered.\footnote{Black hole solutions in other gravity models exhibiting
Lifshitz scaling have been considered 
in \cite{Taylor:2008tg,Azeyanagi:2009pr,Pang:2009ad}.} 
These Lifshitz black holes carry a charge that couples 
to the two-form field strength $F_{(2)}$, but they are in many respects more 
analogous to AdS-Schwarzschild black holes than charged 
AdS-Reissner-Nordstr\"om black holes. In particular, since the black hole 
charge cannot be varied independently of the black hole area, there is only a 
one-parameter family of black hole solutions for a given value of $z$. 

We will see below that without additional ingredients the underlying Lifshitz 
symmetry prevents the system from undergoing phase transitions.
We therefore extend the system to include a Maxwell field, with 
field strength $\mathcal{F}_{(2)}$, and a scalar field $\psi$ that is charged under 
the new gauge field but neutral under the original Lifshitz fields $F_{(2)}$ and 
$H_{(3)}$. This turns out to be sufficient in order to observe a superconducting 
phase transition characterized by Lifshitz scaling with $z>1$. In our modified 
theory it is the Maxwell field $\mathcal{F}_{(2)}$ that corresponds to physical 
electromagnetism while the original Lifshitz gauge fields are viewed as an 
auxiliary construction, whose only role is to modify the asymptotic symmetry of 
the geometry from AdS to Lifshitz. 

The plan of the rest of the paper is as follows. We begin in Section~\ref{bhgeometry} 
with a brief review of Lifshitz black holes in the model (\ref{actionform}). We extend
previous treatments by considering the global geometry, including the black hole 
interior. In those cases where an exact solution is known, we find a null 
curvature singularity at $r=0$ and a Carter-Penrose diagram as shown in 
Figure~\ref{fig:penrose}. We show that a null singularity is generic for black holes 
in this model. In Section~\ref{chargedbhs} we generalize the
model to include an additional Maxwell field. We exhibit a family of exact solutions
at $z=4$, which describe Lifshitz black holes that are charged under the new
Maxwell field and obtain numerical solutions for charged black holes at other 
values of $z>1$. These
black holes can be viewed as the Lifshitz analog of AdS-Reissner-Nordstr\"om 
black holes. In Section~\ref{scalarfield} we further extend the model by adding
a charged scalar field and look for the black hole instability that signals the 
onset of superconductivity. Finally, we wrap up with some final remarks in 
Section~\ref{conclusions}.

\section{Lifshitz black holes revisited}
\label{bhgeometry}
The action (\ref{actionform}) is known to have spherically symmetric 
static black hole solutions of the form
\begin{equation}
ds^{2}  =L^{2}\left(  -r^{2z}f\left(  r\right)  ^{2}dt^{2} +\frac{g\left(
r\right)  ^{2}}{r^{2}}dr^{2} +r^{2}\left( d\theta^{2}+\chi(\theta)^{2}
d\varphi^{2}\right)  \right) ,
\label{trmetric}
\end{equation}
with
\begin{equation}
\chi(\theta)=\left\{
\begin{array}
[c]{ccl}%
\sin\theta & \quad\text{if}\quad & k = 1, \nonumber\\
\theta & \quad\text{if}\quad & k = 0,\\
\sinh\theta & \quad\text{if}\quad & k = -1,
\end{array}
\right.
\end{equation}
where $k=+1,0,-1$ corresponds to a spherical, flat, and hyperbolic horizon
respectively. An asymptotically Lifshitz black hole with a non-degenerate horizon
has $f(r),\,g(r) \rightarrow 1$ as $r\rightarrow\infty$ and a simple zero of both $f(r)^2$ 
and $g(r)^{-2}$ at the horizon $r=r_0$. 

\psfrag{y}[c]{$r=\infty$}
\psfrag{x}[c]{$r=0$}
\FIGURE{\epsfig{file=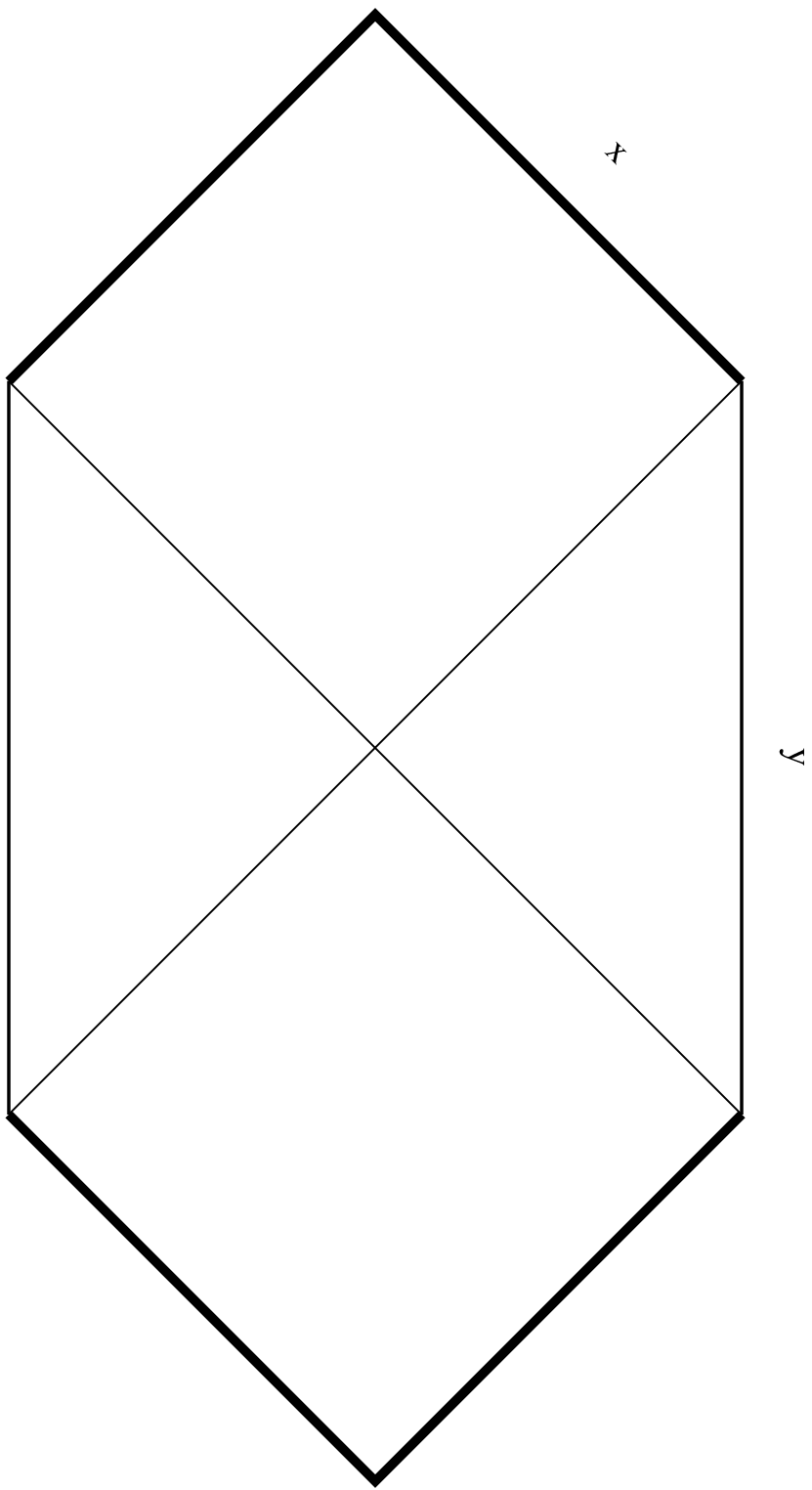,width=4.5cm}
\caption{Conformal diagram for a Lifshitz black hole. The thick lines denote null 
curvature singularities.}
\label{fig:penrose}}

\subsection{Exact solutions and global extensions}
\label{exactsols}
Several families of numerical solutions of this 
general form have been found \cite{Danielsson:2009gi,Mann:2009yx,Bertoldi:2009vn} 
along with a couple of isolated exact black hole solutions. These are a $z=2$ 
topological black hole with a hyperbolic horizon \cite{Mann:2009yx}, 
\begin{equation}
\label{mann}
f(r)=\frac{1}{g(r)}=\sqrt{1-\frac{1}{2r^2}} ,
\end{equation}
and a $z=4$ black hole \cite{Bertoldi:2009vn}, 
\begin{equation}
\label{bertoldi}
f(r)=\frac{1}{g(r)}=\sqrt{1+\frac{k}{10r^2}-\frac{3k^2}{400r^4}} ,
\end{equation}
With $k=\pm 1$. The $k=+1$ case has a spherical horizon while $k=-1$ corresponds 
to a topological black hole with a hyperbolic horizon.
When we extend the system to include an additional Maxwell gauge field (see
Section~\ref{chargedbhs} below) we will see that the exact $z=4$ solution 
(\ref{bertoldi}) is a special case of a one-parameter family
of exact solutions in the extended theory. 

We start our discussion by working out the global geometry of the exact 
$z=2$ topological black hole (\ref{mann}) and constructing the associated conformal 
diagram. In order to study the interior geometry of a black hole one looks for
coordinates that are non-singular at the black hole horizon and allow the
solution to be extended into the black hole. 
The first step is to transform to a tortoise coordinate $r_{*}$ for which
\begin{equation}
ds^{2}=L^{2}\left(  r^{4}\left( 1-\frac{1}{2r^{2}}\right) (-dt^{2}+dr_{*}^{2})
+r^{2}(d\chi^{2}+\sinh^2\chi d\varphi^{2})\right) .\label{tortoisemetric}
\end{equation}
Here $r$ is to be viewed as a function of $r_{*}$ determined via
\begin{equation}
\frac{dr}{dr_{*}} = r^{3}\left( 1-\frac{1}{2r^{2}}\right) ,
\end{equation}
which integrates to
\begin{equation}
r = \frac{1}{\sqrt{2}\sqrt{1-\exp(r_{*}-r_{*}^{\infty})}}.
\end{equation}
As usual, the tortoise coordinate $r_{*}$ goes to $-\infty$ at the horizon but
note that it goes to a finite value in the $r\rightarrow\infty$ asymptotic region.
This is easily seen to be a generic feature of asymptotically Lifshitz black holes
for any $z>1$. 

Next we form the null combinations $v=t+r_{*}$, $u=t-r_{*}$ and perform a
conformal reparametrization,
\begin{eqnarray}
v&\rightarrow &V=\exp{\left[ \frac{1}{2}(v-r_{*}^{\infty})\right] } , \nonumber \\
u&\rightarrow &U=-\exp{\left[- \frac{1}{2}(u+r_{*}^{\infty})\right] } ,\qquad
\label{mannconfrep}
\end{eqnarray}
which brings the metric into the form
\begin{equation}
ds^{2}=L^{2}\left(  -4 r^{4}dU\,dV +r^{2}(d\theta^{2}+\sinh^{2}\theta d\varphi^{2})\right) .
\end{equation}
In these coordinates the geometry is manifestly nonsingular at the event horizon
and the solution can be extended to $r<1/\sqrt{2}$. 

In general, when extending a black hole solution through a non-degenerate horizon 
the required conformal reparametrization is of the form 
\begin{eqnarray}
v&\rightarrow &V=\alpha \exp{(\kappa v)}, \nonumber \\
u&\rightarrow &U=-\alpha \exp{(-\kappa u)} ,\qquad
\label{confrep}
\end{eqnarray}
where $\alpha$ is an arbitrary constant and $\kappa$ is the surface gravity of the black hole
in question. The surface gravity determines the Hawking temperature, 
$T_H=\frac{\kappa}{2\pi}$, and it is easily checked by other means that 
$T_H=\frac{1}{4\pi}$ is the correct value for the $z=2$ black hole in
(\ref{mann}).  We have chosen the value of the constant $\alpha$ in (\ref{mannconfrep}) such
that $UV\rightarrow -1$ in the $r\rightarrow\infty$ asymptotic limit.

The $z=2$ topological black hole has a curvature singularity at $r=0$, which can for 
example be seen by
computing the Ricci scalar,\footnote{The fixed point geometry (\ref{lmetric}) is
also singular at $r=0$ but in a relatively mild way. All scalar invariants
constructed from the Riemann tensor are finite but tidal effects between
neighboring null geodesics diverge as $r\rightarrow0$. 
See f.ex. \cite{Hartnoll:2009sz}.}
\begin{equation}
R=\frac{1}{r^{2}L^{2}}\left( 1-22\,r^{2}\right) .
\end{equation}
It follows from 
\begin{equation}
-UV=1-\frac{1}{2r^{2}},
\end{equation}
that $r\rightarrow\infty$ corresponds to $UV\rightarrow -1$ and 
$r \rightarrow 0$ to $UV\rightarrow \infty$.  
The Carter-Penrose conformal diagram\footnote{We refer to Figure~\ref{fig:penrose} 
as a conformal diagram but it should be kept in mind that the boundary at $r\rightarrow\infty$ 
is in fact not conformally flat due to the asymmetric scaling of the spatial and time directions 
in asymptotically Lifshitz spacetime.} in Figure~\ref{fig:penrose} is then obtained by writing 
\begin{equation}
V =\tan{\frac{\pi P}{2}}, \qquad U=\tan{\frac{\pi Q}{2}} .
\end{equation}

The $z=4$ case can be dealt with in a similar manner. There is a null
curvature singularity at $r=0$ and the 

The tortoise coordinate for a $z=4$ black hole is given by 
\begin{equation}
r_*-r_*^\infty
=\frac{1}{2b_1(b_1+b_2)} \log\left(1-\frac{b_1}{r^2}\right)
+\frac{1}{2b_2(b_1+b_2)}\log\left(1+\frac{b_2}{r^2}\right),
\label{z4tortoise}
\end{equation}
with $b_1=\frac{1}{20}(2\vert k\vert-k)$ and $b_2=\frac{1}{20}(2\vert k\vert+k)$.
The change of coordinates to 
\begin{equation}
V=\exp\left[b_1(b_1+b_2)(v-r_*^\infty)\right], \qquad
U=-\exp\left[-b_1(b_1+b_2)(u+r_*^\infty)\right],
\label{z4uv}
\end{equation}
renders the metric nonsingular at the horizon,
\begin{equation}
ds^2=L^{2}\left(  -\frac{r^8}{\kappa^2}
\left(1+\frac{b_2}{r^2}\right)^{1-\frac{b_1}{b_2}}dU\,dV 
+r^{2}(d\theta^{2}+\chi^{2}(\theta) d\varphi^{2})\right) .
\label{z4regularmetric}
\end{equation}
The Hawking temperature of these $z=4$ black holes is 
\begin{equation}
T_H=\frac{1}{200\pi}(2-k),
\end{equation}
and the Ricci scalar is given by
\begin{equation}
R=\frac{1}{L^{2}}\left(\frac{9k^2}{200r^4}-\frac{3k}{5r^2}-54\right) ,
\end{equation}
There is a null curvature singularity at $r\rightarrow 0$ and the 
conformal diagram is that of Figure~\ref{fig:penrose}. 
In Section~\ref{globalgeom} below, we present a general argument that
this conformal diagram applies to all black hole solutions of the system of 
equations (\ref{formeq}) - (\ref{einsteineqs}).

\subsection{Lifshitz black holes at general $z>1$}
The exact solutions offer a glimpse at the parameter space of Lifshitz
black holes at isolated points. More generic solutions can be obtained
numerically. We are interested in the global geometry including the
black hole interior. For this we take our cue from the extension
of the exact solution described above and write the metric in the form
\begin{equation}
ds^{2}  = L^{2}\left[ -e^{2\rho(v,u)} dv\,du
+e^{-2\phi(v,u)}\left( d\theta^{2}+\chi(\theta)^{2}d\varphi^{2}\right)  \right] .
\label{confmetric}
\end{equation}
The notation is descended from two-dimensional gravity and allows
the field equations to be witten in a relatively economical way. With this 
ansatz the metric is characterized by two field variables, $\rho$ and $\phi$, 
which determine the local scale of the $v,u$ plane and the scale of the 
transverse two-manifold respectively. The null coordinates $v,u$ are
related to the $t,r$ coordinates in (\ref{trmetric}) via $v=t+r_*$, $u=t-r_*$,
with the tortoise coordinate $r_*$ given by
\begin{equation}
r_*-r_*^\infty
=-\int_r^\infty \frac{d\tilde{r}}{\tilde{r}^{z+1}}\frac{g(\tilde{r})}{f(\tilde{r})} .
\label{tort}
\end{equation}
The Lifshitz fixed point geometry (\ref{lmetric}) has $f(r)=g(r)=1$ and in that case
\begin{equation}
r_*=r_*^\infty - \frac{1}{z\, r^z} .
\label{lifshitztort}
\end{equation}
More generally, we see from (\ref{tort}) that $r_*$ goes to a finite 
value asymptotically for any Lifshitz spacetime, for
which $f(r),g(r)\rightarrow 1$ as $r\rightarrow\infty$.

\subsubsection{The equations of motion}
With the following ansatz for the form fields,
\begin{eqnarray}
F_{\left(  2\right)  }  & =&L\, f_{vu}(v,u)\,dv\wedge du,\\
H_{\left(  3\right)  }  & =&
L^2 \left(h_v(v,u) dv-h_u(v,u) du\right)
\wedge d\theta \wedge \chi(\theta)d\varphi ,
\end{eqnarray}
their field equations (\ref{formeq}) become
\begin{eqnarray}
-\partial_v\left(e^{-2\rho-2\phi}f_{vu}\right)&=&\sqrt{\frac{z}{2}}\,h_v, \\
\partial_u\left(e^{-2\rho-2\phi}f_{vu}\right)&=&\sqrt{\frac{z}{2}}\,h_u, \\
-\partial_v\left(e^{2\phi}h_u\right) + \partial_u\left(e^{2\phi}h_v\right)
&=&\sqrt{2z}\, f_{vu}.
\end{eqnarray}
These can in turn be re-expressed as a single second-order equation,
\begin{equation}
\partial_v\partial_u\tilde{f}+\partial_v\phi\partial_u\tilde{f}
+\partial_u\phi\partial_v\tilde{f}+\frac{z}{2}e^{2\rho}\tilde{f} =0,
\end{equation}
where we have defined
\begin{equation}
\tilde{f}\equiv e^{-2\rho-2\phi}f_{vu}.
\end{equation}
The equations of motion for $\phi$ and $\rho$ are obtained from (\ref{einsteineqs}), 
\begin{eqnarray}
-\partial_v\partial_u\phi+2\partial_v\phi\partial_u\phi
+\frac{k}{4}e^{2\rho+2\phi}+\frac{(z^2{+}z{+}4)}{8}e^{2\rho}
-\frac{1}{4}\tilde{f}^2e^{2\rho+4\phi}&=&0, \\
-\partial_v\partial_u\rho+\partial_v\phi\partial_u\phi+\frac{k}{4}e^{2\rho+2\phi}
-\frac{1}{2}\tilde{f}^2e^{2\rho+4\phi}+\frac{1}{2z}e^{4\phi}\partial_v\tilde{f}\partial_u\tilde{f}&=&0.
\end{eqnarray}
The conformal reparametrization (\ref{confrep}) will render the metric 
non-degerate at the horizon. The form of the field equations remains the same
in the $V,U$ coordinate system with a transformed conformal factor,
\begin{equation}
e^{2\rho(v,u)}=(-\kappa^2UV) \, e^{2\rho(V,U)}.
\label{rhotransformed}
\end{equation}
We now adapt a simple numerical method, which was originally developed for the study of 
two-dimensional black holes \cite{Birnir:1992by}, to the case at hand. We introduce 
a new spatial variable,
\begin{equation}
s\equiv -UV ,
\label{svariable}
\end{equation}  
and restrict our attention to static configurations. The field equations become
\begin{eqnarray}
0&=&s\tilde{f}''+\tilde{f}'+2s\phi'\tilde{f}'-\frac{z}{2}\tilde{f}e^{2\rho}, \label{feqins}\\
0&=&s\phi''+\phi'-2s\phi'^2+\frac{k}{4}e^{2\rho+2\phi}+\frac{(z^2{+}z{+}4)}{8}e^{2\rho}
-\frac{1}{4}\tilde{f}^2e^{2\rho+4\phi}, \label{phieqins}\\
0&=&s\rho''+\rho'-s\phi'^2+\frac{k}{4}e^{2\rho+2\phi}
-\frac{1}{2}\tilde{f}^2e^{2\rho+4\phi}-\frac{s}{2z}e^{4\phi}\tilde{f}'^2,\label{rhoeqins}
\end{eqnarray}
where primes denote derivatives with respect to $s$.

The Lifshitz geometry (\ref{lmetric}) is given by 
\begin{equation}
\tilde{f}=\sqrt{\frac{z(z-1)}{2}}r^2, \quad e^{-\phi}=r, \quad
\kappa^2 s\,e^{2\rho(s)}=r^{2z},
\label{lifshitzgeom}
\end{equation}
with $r(s)$ obtained from (\ref{lifshitztort}) and (\ref{svariable}).
The horizon at $r=0$ is singular and as a result $\kappa$ can take 
any value in the Lifshitz geometry. The $z=2$ topological black hole has 
$\tilde{f}=r^2$ with $r(s)=1/\sqrt{2(1-s)}$, while the $z=4$ black holes turn out to
have $\tilde{f}=\sqrt{6}(r^2+\frac{k}{20})$, with $r(s)$ obtained via 
$\frac{dr}{dr_*}=r^5(1+\frac{k}{10r^2}-\frac{3}{400r^4})$ and (\ref{svariable}).

\subsubsection{Numerical solutions}
The event horizon is at $s=0$ and $s$ is negative inside the black hole. With the 
assumption of a regular horizon we can read off the following relations among
initial values at $s=0$,
\begin{eqnarray}
\phi'(0)&=&\frac{1}{4}\left(-ke^{2\phi(0)}-\frac{(z^2{+}z{+}4)}{2}
+\tilde{f}(0)^2e^{4\phi(0)}\right)e^{2\rho(0)}, \\
\tilde{f}'(0)&=&\frac{z}{2}\,\tilde{f}(0)\,e^{2\rho(0)}, \\
\rho'(0)&=&\frac{1}{2}\left(-\frac{k}{2}e^{2\phi(0)}
+\tilde{f}(0)^2e^{4\phi(0)}\right)e^{2\rho(0)}.
\end{eqnarray}
These initial values can now be used to start a numerical integration of the system
(\ref{feqins})-(\ref{rhoeqins}) from near $s=0$,
either outwards towards $s>0$ or into the black hole interior at 
$s<0$. Inequivalent solutions are parametrized by $\phi(0)$ and $\tilde{f}(0)$. 
The $\phi(0)$ initial value gives the value of the area 
coordinate at the horizon through the relation $r=e^{-\phi}$, and $\tilde{f}(0)$
determines the magnitude of the radial two-form field at the horizon. A shift
of $\rho(0)$ amounts to a global rescaling of the $s$ coordinate and does not
affect the geometry.

As discussed in \cite{Danielsson:2009gi}, $\phi(0)$ and $\tilde{f}(0)$ cannot be
varied independently while preserving the asymptotically Lifshitz character of
the geometry. In the $z=2$ case considered in that paper, there is a zero mode
of the linearized system of equations near the Lifshitz fixed point, which must be
set to zero for the system to approach the fixed point geometry (\ref{lmetric}) as
$r\rightarrow\infty$. This requires a fine-tuning of parameters which uniquely
determines $\tilde{f}(0)$ in terms of $\phi(0)$, or vice versa. In \cite{Bertoldi:2009vn}
it was shown that at $z>2$ the zero mode is replaced by a power-law growing 
mode which must be set to zero to obtain an asymptotically Lifshitz geometry.
The question was more subtle at $z<2$ where instead there is a mode with a weak
power-law fall-off but a finite energy argument \cite{Bertoldi:2009vn,Ross:2009ar} 
may be employed to 
conclude that this mode must also be set to zero.\footnote{By the energy 
argument of \cite{Bertoldi:2009vn}, a second mode of the linearized system
should also be set to zero in the $z<2$ case, leading to at best a discrete 
spectrum of Lifshitz black holes at $z<2$. This condition is too strong, indicating
a problem with the background subtraction used in \cite{Bertoldi:2009vn}. An 
alternative definition of the energy of asymptotically Lifshitz solutions, based on 
boundary counterterms, recently appeared in \cite{Ross:2009ar} for which only a 
single mode needs to be fine-tuned away in order to have finite energy at $z<2$.} 

With the variables that we are using here there is a simple criterion to identify the 
subset of initial values $\phi(0)$ and $\tilde{f}(0)$ that lead to asymptotically Lifshitz 
black holes for any value of $z>1$.
We already learned from (\ref{tort}) that the numerical integration towards $s>0$, 
{\it i.e.} towards the asymptotic region, will terminate at a finite value $s=s_\infty$.  
We also know that $e^{2\rho(v,u)}\rightarrow r^{2z}$, as $r\rightarrow\infty$ for
asymptotically Lifshitz geometry. For a given $\phi(0)$ we tune 
$\tilde{f}(0)$ until the combination $\rho(s)+z\phi(s)$ goes to a finite value in 
the limit $s\rightarrow s_\infty$. Furthermore, through the transformation rule 
(\ref{rhotransformed}) for the conformal factor we can read
off the Hawking temperature of the black hole,
\begin{equation}
T_H=\frac{\kappa}{2\pi}=\frac{1}{2\pi\sqrt{s_\infty}}
\lim_{s\rightarrow s_\infty} e^{-\rho(s)-z\phi(s)} ,
\label{thawking}
\end{equation}
once we have fine-tuned the initial data to give a finite limit. This method works
the same way for both $z>2$ and $z<2$. Although the unwanted mode is
slowly decaying as a function of $r$ when $r\rightarrow\infty$ in the $z<2$ 
case, it has a growing amplitude as a function of $s$
due to the rapid growth of $\frac{dr}{dr_*}$ as $s\rightarrow s_\infty$.

\FIGURE{\epsfig{file=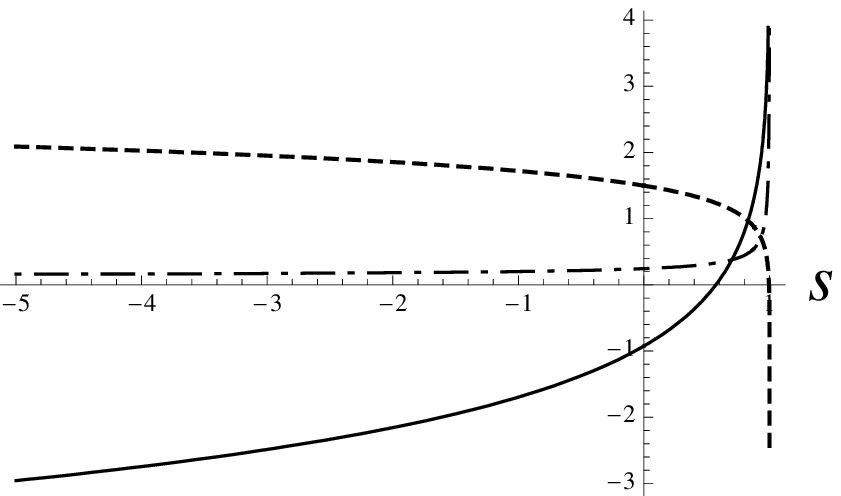,width=7cm}
\caption{Numerical solution for a $z=4$ Lifshitz black hole with $\rho$ solid, $\phi$ 
dashed, and $f$ dot-dashed. The horizon is at $s=0$ and $r\rightarrow \infty$ 
corresponds to $s\rightarrow 1$.}
\label{fig:numericbh}}

A numerical solution for both the exterior and interior geometry can now be obtained 
by separately integrating towards positive and negative $s$, using the 
same fine-tuned initial data, and then patching the two solutions together
across $s=0$. Figure~\ref{fig:numericbh} shows a numerical black hole solution
at $z=4$ obtained in this manner. 
The curvature singularity is at $s\rightarrow -\infty$ in these coordinates and 
the numerical evaluation will break down before it is reached. 

\subsection{Global geometry of generic Lifshitz black holes}
\label{globalgeom}
We now give an argument that the conformal diagram in Figure~\ref{fig:penrose} 
applies to a generic Lifshitz black hole solution that has a single non-degerate 
horizon and is obtained from the action 
(\ref{actionform}), for some value of $z>1$. 
To do that, one examines the asymptotic behavior of solutions to the system of
equations (\ref{feqins}) - (\ref{rhoeqins}) near the singularity, {\it i.e.} as 
$s\rightarrow-\infty$. The structure of the equations restricts the asymptotic
behavior to one of two types. The first type has
\begin{equation}
\label{type1sol}
\textrm{I:}\quad e^{\phi}=\left(  -s\right)^\alpha e^{\phi_0}+\ldots ,\quad
e^{\rho}=\left(  -s\right)^{-\alpha-1/2}e^{\rho_0}+\ldots ,\quad 
\tilde{f}=\left(  -s\right)^{-2\alpha}f_0+\ldots ,
\end{equation}
where $\alpha$, $\phi_0$, $\rho_0$, and $f_0$ are non-universal constants.
The assumption that $r\rightarrow 0$ as $s\rightarrow-\infty$ amounts to the
requirement $\alpha>0$ but otherwise these constants are unrestricted 
{\it a priori}.\footnote{We note that our argument assumes that 
$r\rightarrow 0$ as $s\rightarrow-\infty$ and does not preclude the existence
of solutions with a smooth inner horzion at which $r\rightarrow r_-$ as 
$s\rightarrow-\infty$ for some finite $r_->0$. On the other hand, none of the
known exact solutions at $z>1$ have such an inner horizon.}

With $\alpha>0$ the last term in (\ref{feqins}), the last two terms in (\ref{phieqins}), 
and the next to last term in (\ref{rhoeqins}) are sub-leading for this type of 
solution. The remaining terms in (\ref{feqins}) cancel automatically at leading
order, while (\ref{phieqins}) and (\ref{rhoeqins}) are satisfied at leading order if
\begin{equation}
\label{type1cond}
\frac{k}{8}e^{2\rho_0+2\phi_0}=-\alpha^2 , \qquad
f_0^2 e^{4\phi_0}=\frac{z}{2}.
\end{equation}
It follows that this type of asymptotic behavior can only be realized in a $k=-1$ 
geometry. It is easily checked that the $z=2$ topological black hole solution
is of this type with $\alpha=\frac{1}{2}$.

The other possible behavior near the singularity is
\begin{eqnarray}
\label{type2sol}
\textrm{II:}&\qquad & e^{\phi}=\left(  -s\right)^\beta e^{\phi_0}+\ldots ,\quad 
e^{\rho}=\left(  -s\right)^{-2\beta-1/2}e^{\rho_0}+\ldots ,
\quad \tilde{f}=f_0+\ldots, \nonumber \\ 
&\qquad & \tilde{f}'=\left(  -s\right)^{-2\beta-1}f_1+\ldots ,
\end{eqnarray}
with $\beta>0$. In this type of solution the
last term in (\ref{feqins}), the fourth and fifth terms in (\ref{phieqins}),
and the fourth term in (\ref{rhoeqins}) are subleading as $s\rightarrow -\infty$.
The remainder of (\ref{feqins}) is automatically satisfied at leading order, while 
(\ref{phieqins}) and (\ref{rhoeqins}) require,
\begin{equation}
\label{type2cond}
\frac{f_0^2}{8}e^{2\rho_0+4\phi_0}=\beta^2 , \qquad
f_1^2=\frac{3z}{4}f_0^2 e^{2\rho_0}.
\end{equation}
This time around there is no restriction on the value of $k$.
The exact $z=4$ black hole solutions fall into this category, with $\beta=\frac{3}{8}$
in the $k=1$ case and $\beta=\frac{1}{8}$ in the $k=-1$ case.

The Ricci scalar diverges in the $s\rightarrow -\infty$ limit. For type~I solutions the
leading behavior is $R=\frac{1}{L^2r^2}+\ldots$, while for type~II solutions we find
$R=\frac{3f_0^2}{L^2r^4}+\ldots$. The curvature singularity at $r\rightarrow 0$ 
is always null. This follows immediately from the fact that 
\begin{equation}
e^{2\rho} \rightarrow 0 , \quad \textrm{as}\quad s\rightarrow -\infty,
\end{equation} for both allowed types of asymptotic behavior. As a result, the
conformal diagram in Figure~\ref{fig:penrose} describes generic asymptotically 
Lifshitz black holes with a single non-degenerate horizon that are obtained 
from the action (\ref{actionform}).

In the notation used in \cite{Danielsson:2009gi}, appropriately continued to inside 
the horizon, a type~I solution has $f\propto r^{1-z}$, $g\propto r^{2}$,
$h\propto1$ and $j\propto r^{-1}$, for small $r$. The corresponding limiting 
behavior of a solution of type~II is $f\propto r^{2-z}$, $g\propto r^{2}$,
$h\propto r^{-2}$ and $j\propto r^{-2}$.

\section{Charged Lifshitz black holes}
\label{chargedbhs}
We wish to extend the notion of a holographic superconductor to 
systems with Lifshitz scaling. In order to do this, we need to introduce some
additional ingredients to the gravity model that we have been considering. 
As it stands, the model has only a single characteristic length scale and 
does not support a superconducting transition, or any other phase transition 
for that matter. If we want to make phase transitions possible, we need to 
introduce a new scale into the problem. This can be seen 
explicitly as follows. The gravitational dual of a holographic superconductor 
is a charged plane-symmetric black hole with hair \cite{Hartnoll:2008kx},
so for this application we work with $k=0$ geometries of the form,
\begin{equation}
ds^{2}  = L^{2}\left[ -e^{2\rho(v,u)} dv\,du
+e^{-2\phi(v,u)}\left( dx^{2}+dy^{2}\right)  \right] .
\label{xymetric}
\end{equation}
The plane-symmetric black hole geometry has a scaling symmetry, which is 
absent for spherical or hyperbolic horizons. The metric is invariant under
a global rescaling of the planar coordinates $x,y\rightarrow e^\alpha x, e^\alpha y$ 
accompanied by a uniform shift $\phi\rightarrow \phi+\alpha$ while keeping $\rho$,
$v$, and $u$ unchanged. On the other hand, such a shift changes the Hawking
temperature (\ref{thawking}). Since different non-zero Hawking temperatures can 
be mapped into each other by scale transformations they are all physically 
equivalent in this system and there cannot be any phase transitions. Another way
to state this is that in a scale invariant theory at finite temperature there is no
other length scale apart from that defined by the temperature itself and therefore
different temperatures must be physically equivalent.

There are different ways to add another length scale to the Lifshitz system. 
We choose to do it by letting our black holes carry an electric charge, which 
couples to  a Maxwell field $\mathcal{F}_{(2)}$. This corresponds to introducing 
a non-zero charge density into the dual field theoretic system \cite{Hartnoll:2008kx}
and adds a term to the action,
\begin{equation}
S_\mathcal{F}=-\frac{1}{2}\int\ast \mathcal{F}_{(2)}\wedge \mathcal{F}_{(2)} ,
\end{equation}
which has the same form as the kinetic term of $F_{(2)}$ and contributes
in same way to the field equations. Introducing a new gauge field may, at first 
sight, appear to be an unnecessary complication given that the model already has
a two-form field strength, $F_{(2)}$, and a three-form field strength, $H_{(3)}$, which
couples to the two-form field and could be viewed as a form of charged matter. 
While such an interpretation could in principle be pursued, we find it more useful to 
consider $F_{(2)}$ and $H_{(3)}$ as auxiliary fields, that only couple to gravity, and 
whose only role is to give rise to gravitational backgrounds with Lifshitz scaling. 
We are then free to separately include the physical electromagnetic field 
$\mathcal{F}_{(2)}$ and introduce 
charged matter that couples to it but not to the Lifshitz form-fields. The manner 
in which we construct our holographic superconductors with Lifshitz scaling is
analogous to the corresponding construction for the $z=1$ case in \cite{Hartnoll:2008kx}.
In fact, in the limit $z\rightarrow 1$, our system reduces to the one considered by those
authors and we have used this to check our numerical algorithm against known results 
for $z=1$.

Writing 
\begin{equation}
\mathcal{F}_{\left(  2\right)  }   =L\, p_{vu}(v,u)\,dv\wedge du,
\end{equation}
the Maxwell equations for $\mathcal{F}_{(2)}$ reduce to
\begin{equation}
\partial_v\left(e^{-2\rho-2\phi}p_{vu}\right)=0=
\partial_u\left(e^{-2\rho-2\phi}p_{vu}\right).
\label{sourcefree}
\end{equation}
The general solution can be written 
\begin{equation}
p_{vu}=Q e^{2\rho+2\phi},
\label{coulombfield}
\end{equation}
and has a simple interpretation as the
Coulomb field of a point charge. Now consider a black hole with charge $Q$ in
the $s$ variable. Since the new gauge field does not couple directly to the original 
Lifshitz gauge fields, the field equation for $\tilde{f}$ (\ref{feqins}) remains unchanged 
while equations (\ref{phieqins}) and (\ref{rhoeqins}) pick up terms involving the 
black hole charge,\footnote{For 
completeness we allow for $k\neq 0$ in the field equations but we are primarily 
interested in the $k=0$ case for the application to holographic superconductors.}
\begin{eqnarray}
0&=&s\phi''+\phi'-2s\phi'^2+\frac{k}{4}e^{2\rho+2\phi}+\frac{(z^2{+}z{+}4)}{8}e^{2\rho}
-\frac{1}{4}(\tilde{f}^2+Q^2)e^{2\rho+4\phi}, \label{Qphieqins}\\
0&=&s\rho''+\rho'-s\phi'^2+\frac{k}{4}e^{2\rho+2\phi}
-\frac{1}{2}(\tilde{f}^2+Q^2)e^{2\rho+4\phi}-\frac{s}{2z}e^{4\phi}\tilde{f}'^2. 
\label{Qrhoeqins}
\end{eqnarray}
The field equations 
can be numerically integrated as before and we will present some numerical 
results below, but before that we present some exact solutions. In the 
first example we recover the well known AdS-Reissner-Nordstr\"om solution 
as a special case with $z=1$ and $\tilde{f}=0$. This provides a check of the
formalism. The remaining examples are new and describe one-parameter 
families of charged Lifshitz black holes at $z=4$.

\subsection{AdS-RN black holes at $z=1$}
\label{adsrn}
The AdS-Reissner-Nordstr\"om solution at $z=1$ describes an electrically 
charged black hole in asymptotically AdS spacetime. In the variables we are 
using it is given by
\begin{equation}
e^{2\rho(r_*)}=r^2+k-\frac{1}{r}\left(
r_h^3+k r_h+\frac{Q^2}{r_h}\right)+\frac{Q^2}{r^2} ,
\end{equation}
with $k=+1,0,-1$ for a spherical, flat, or hyperbolic horizon at $r=r_h$.
The relationship between the area and tortoise coordinates is
\begin{equation}
\frac{dr}{dr_*}=r^2+k-\frac{1}{r}\left(
r_h^3+k r_h+\frac{Q^2}{r_h}\right)+\frac{Q^2}{r^2} ,
\end{equation}
and the Lifshitz gauge field $\tilde{f}$ is everywhere vanishing. It is easily
checked that 
\begin{equation}
e^{2\rho(s)}=\frac{1}{\kappa^2 s}e^{2\rho(r_*)}, \qquad
e^{-\phi(s)}= r ,
\end{equation}
with $s=e^{2\kappa(r_*-r_*^\infty)}$ is a solution of equations (\ref{Qphieqins}) and
(\ref{Qrhoeqins}) with $\tilde{f}=0$.

The Hawking temperature,
\begin{equation}
T_H=\frac{1}{4\pi}\left(3r_h+\frac{k}{r_h}-\frac{Q^2}{r_h^3} \right) ,
\label{RNTHawking}
\end{equation}
goes to zero as the charge approaches the extremal value for a given horizon
area,
\begin{equation}
T_H\rightarrow 0  \quad \textrm{as} \quad 
Q^2 \rightarrow Q_\textrm{ext}^2 = 3r_h^4+k r_h^2 \,,
\end{equation}
as shown in Figure~\ref{fig:TQcurves}.

\subsection{Exact solutions for charged black holes at $z=4$}
Numerical work is required in order to explore the full parameter range of
asymptotically Lifshitz black hole solutions at $z>1$ but exact solutions are useful,
even if they only apply in special cases. We have found a family of exact
charged black hole solutions for $z=4$, which generalize the isolated $z=4$ solutions
discussed in Section~\ref{exactsols}. In the notation of equations (\ref{mann}) -
(\ref{bertoldi}) the metric is given by 
\begin{equation}
\label{Qexact}
f(r)=\frac{1}{g(r)}=\sqrt{1+\frac{k}{10r^2}-\frac{3k^2}{400r^4}-\frac{Q^2}{2r^4}} ,
\end{equation}
with $k=+1,0,-1$. The metric reduces to the $z=4$ solutions of Section~\ref{exactsols} 
when $Q=0$ and $k=\pm 1$ while the $Q=0$, $k=0$ case reduces to the $z=4$ Lifshitz 
fixed point geometry. 

The tortoise coordinate is given by equation (\ref{z4tortoise}) with
\begin{eqnarray}
b_1&=&\sqrt{\frac{k^2}{100}+\frac{Q^2}{2}}-\frac{k}{20} , \\
b_2&=&\sqrt{\frac{k^2}{100}+\frac{Q^2}{2}}+\frac{k}{20} .
\end{eqnarray}
The change to $U,V$ coordinates in (\ref{z4uv}) leads to a metric of the form
(\ref{z4regularmetric}), with the new values of $b_1$ and $b_2$, which is 
nonsingular at the horizon. 
The Lifshitz gauge field can be obtained by inserting the $\rho(s)$ 
and $\phi(s)$ read off from (\ref{Qexact}) into (\ref{Qphieqins}) and solving for 
$\tilde{f}(s)$. One finds that $\tilde{f}=\sqrt{6}\left(r^2+\frac{k}{20} \right)$
is independent of the black hole charge $Q$.

The Hawking temperature is 
$T_H=\frac{\kappa}{2\pi}$ with
\begin{equation}
\kappa= Q^2+\frac{k^2}{50}-\frac{k}{10}\sqrt{\frac{k^2}{100}+\frac{Q^2}{2}}.
\end{equation}
The $k=0$ metric has $b_1=b_2$ and takes a particularly simple form in $U,V$
coordinates,
\begin{equation}
ds^2=L^{2}\left(  -\frac{r^8}{Q^2}dU\,dV 
+r^{2}(d\theta^{2}+\theta^2 d\varphi^{2})\right) .
\end{equation}
In this case the horizon is at $r=\frac{Q}{\sqrt{2}}$ and the Hawking temperature 
is $T_H=\frac{Q^2}{2\pi}$, both of which go to zero in the $Q\rightarrow 0$ limit.

Returning to the more general exact solution in (\ref{Qexact}), the Ricci scalar is given by 
\begin{equation}
R=\frac{1}{L^{2}}\left(\frac{3Q^2}{r^4}+\frac{9k^2}{200r^4}-\frac{3k}{5r^2}-54\right) ,
\end{equation}
and is singular at $r\rightarrow 0$. The curvature singularity is null and 
the conformal diagram remains the same as in Figure~\ref{fig:penrose}.
In particular, unlike ordinary Reissner-Nordstr\"om black holes, these charged 
Lifshitz black holes do not have an inner horizon away from the singularity at
$r\rightarrow 0$. The general analysis from Section~\ref{globalgeom} of the asymptotic 
behavior near the singularity can be extended to the case of charged Lifshitz black 
holes. They exhibit type~II behavior with the replacement 
$f_0^2 \rightarrow f_0^2 +Q^2$ in equation (\ref{type2sol}).

\subsection{Numerical charged black hole solutions}
The field equations with the additional Maxwell field included were presented 
earlier in equations (\ref{feqins}), (\ref{Qphieqins}), and (\ref{Qrhoeqins}). 
The initial values giving a smooth horizon at $s=0$ are modified as follows 
by a non-vanishing black hole charge,
\begin{eqnarray}
\phi'(0)&=&\frac{1}{4}\left(-ke^{2\phi(0)}-\frac{(z^2{+}z{+}4)}{2}
+(\tilde{f}(0)^2+Q^2)e^{4\phi(0)}\right)e^{2\rho(0)}, \\
\rho'(0)&=&\frac{1}{2}\left(-\frac{k}{2}e^{2\phi(0)}
+(\tilde{f}(0)^2+Q^2)e^{4\phi(0)}\right)e^{2\rho(0)},\\
\tilde{f}'(0)&=&\frac{z}{2}\,\tilde{f}(0)\,e^{2\rho(0)}.
\end{eqnarray}
For any given value of $z>1$, there is a two-parameter family of asymptotically
Lifshitz black hole solutions. If we take the charge $Q$ and $\phi(0)$ as independent
parameters then $\tilde{f}(0)$ needs to be fine-tuned in order to obtain
the correct asymptotic behavior as $s\rightarrow s_\infty$. Changing the 
value of $\rho(0)$ amounts to a global rescaling of the $s$ coordinate and does not
change the geometry.  

\FIGURE{\epsfig{file=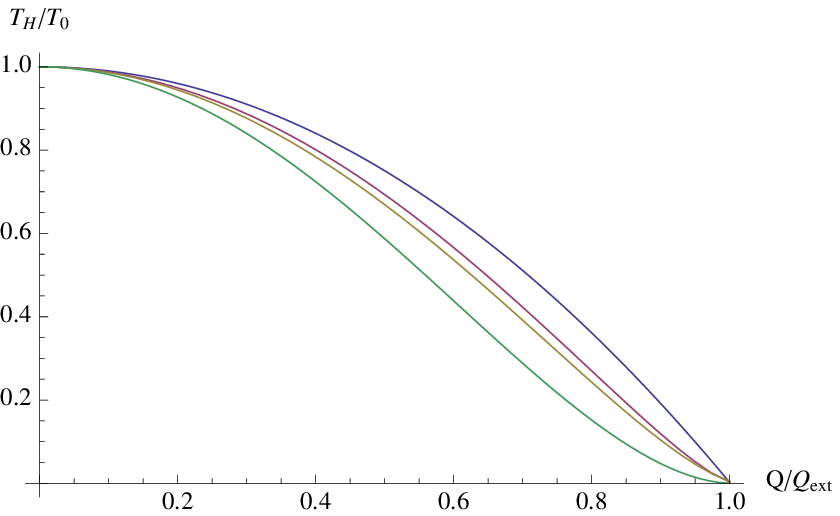,width=7cm}
\caption{The Hawking temperature of charged black holes with $r_h{=}1$, $k{=}{+}1$ for 
$z= 4, 2, \frac{3}{2}, 1$ from bottom to top. The temperature is normalized to the 
temperature of the corresponding uncharged black hole while the charge is normalized 
to the corresponding extremal charge. The curve for $z{=}1$ is plotted from the analytic 
expression (\ref{RNTHawking}) while the $z{>}1$ curves are obtained from numerical solutions.}
\label{fig:TQcurves}}

The numerical integration proceeds in the same way as before and plots of numerical
solutions are qualitatively similar to Figure~\ref{fig:numericbh}. The existence of a new
length scale in the system can be seen in the Hawking temperature obtained from 
(\ref{thawking}) for black holes with different $Q$ keeping the area coordinate at 
the horizon fixed. Figure~\ref{fig:TQcurves} shows the black hole 
temperature as a function of $Q$ for several values of $z$. All the black holes have
$\phi(0)=0$. In order to facilitate comparison, the temperature of all the black
holes for a given $z$ is normalized to the
Hawking temperature of a $Q=0$ black hole with the same value of $z$, and similarly the
charge is normalized to the extremal charge of black holes with the given value of $z$.

\section{Coupling to charged matter}
\label{scalarfield}
The final ingredient is a scalar field $\psi$, which is charged under the new gauge 
field $\mathcal{F}_{(2)}$ but neutral under the original Lifshitz fields $F_{(2)}$ and 
$H_{(3)}$. This allows our black holes to grow scalar hair, and, under certain conditions, 
the black hole hair corresponds to the condensation of a charged operator at low 
temperatures in the dual field theory. The overall picture is quite similar to the one 
obtained for charged AdS black holes in \cite{Hartnoll:2008kx} and extends the notion 
of a holographic superconductor to systems that exhibit Lifshitz scaling with $z>1$.

\subsection{Charged scalar field}
The scalar field action is given by
\begin{equation}
\label{scalaraction}
S_\psi =-\frac{1}{2} \int d^4x \sqrt{-g}
\left(g^{\mu\nu}(\partial_\mu\psi^*+iq\mathcal{A}_\mu\psi^*)
(\partial_\nu\psi-iq\mathcal{A}_\nu\psi)+m^2 \psi^*\psi\right),
\end{equation}
where $\mathcal{A}_\mu$ are the components of a one-form potential for the two-form field 
strength $\mathcal{F}_{(2)}$ and $q$ is the charge of $\psi$. There is an analog in this 
system of the Breitenlohner-Freedman bound \cite{Breitenlohner:1982jf} on the allowed 
mass of the scalar field,
\begin{equation}
\label{BFbound}
L^2\,m^2 > - \frac{(z+2)^2}{4} .
\end{equation}
The requirement that the corresponding Euclidean action be finite places restrictions
on the asymptotic behavior of the scalar field as $r\rightarrow \infty$. This was worked
out in detail for the $z=2$ case in \cite{Kachru:2008yh},\footnote{The discussion in 
\cite{Kachru:2008yh} was in turn based on earlier work on symmetry breaking in the
context of the AdS/CFT correspondence \cite{Klebanov:1999tb}.} and, since the analysis 
carries over to general $z>1$ in a straightforward way, we merely summarize some 
results without going into details. 

The scalar field equation obtained from (\ref{scalaraction}) has two independent 
solutions $\psi(x^\mu) =c_+ \psi_+(x^\mu) + c_- \psi_-(x^\mu)$, that have the asymptotic
form
\begin{equation}
\label{solutions}
\psi_\pm(x^\mu) \rightarrow r^{-\Delta_\pm} \tilde{\psi}_\pm(\tau,\theta,\varphi) + \ldots ,
\end{equation}
at large $r$, where
\begin{equation}
\label{dimensions}
\Delta_\pm = \frac{z+2}{2} \pm \sqrt{\left( \frac{z+2}{2} \right)^2+m^2L^2} .
\end{equation}
If the mass squared satisfies 
\begin{equation}
L^2\,m^2 > 1- \left(\frac{z+2}{2}\right)^2 ,
\end{equation}
then only $\psi_+$ falls off sufficiently rapidly as $r\rightarrow\infty$ and the scalar
field has the boundary condition
\begin{equation}
\label{plussolution}
\psi(x^\mu) \rightarrow r^{-\Delta_+} \left(\tilde{\psi}_+(\tau,\theta,\varphi) 
+  O(\frac{1}{r^2})\right).
\end{equation}
The scalar field is then dual to an operator of dimension 
$\Delta_+>\frac{z}{2}+2$. 

If, on the other hand, the mass squared is in the range
\begin{equation}
1- \left(\frac{z+2}{2}\right)^2>L^2\,m^2 >- \left(\frac{z+2}{2}\right)^2 ,
\end{equation}
then both $\psi_+$ and $\psi_-$ have sufficiently rapid falloff and there is a choice
of two different quantizations for the scalar field. In one case the scalar field is
asymptotic to $\psi_+$ and dual to an operator of dimension $\Delta_+$ with
$\frac{z}{2}+1<\Delta_+< \frac{z}{2}+2$. In the other case the scalar field is 
asymptotic to $\psi_-$ and dual to an operator of dimension $\Delta_-$ with
$\frac{z}{2}<\Delta_-< \frac{z}{2}+1$. 

In our numerical calculations we set the scalar mass squared to 
\begin{equation}
L^2m^2 =\frac{1}{4}- \left(\frac{z+2}{2}\right)^2,
\label{mass2}
\end{equation}
which is inside the range where there is a choice of two boundary theories.
This value leads to convenient values for the operator dimensions,
$\Delta_\pm=\frac{z+2}{2}\pm\frac{1}{2}$. Non-linear descendants of the leading 
scalar field modes are suppressed by $O(\frac{1}{r^2})$ at $r\rightarrow \infty$ and
this choice of mass squared ensures that the first descendant of $\psi_-$ falls off
faster than $\psi_+$. 

\subsection{Lifshitz black holes with scalar hair}
We will look for static spherically symmetric black hole 
solutions using a metric of the form (\ref{confmetric}). We make the gauge choice,
\begin{equation}
\mathcal{A}= L(a_v dv+ a_u du) ,
\end{equation}
with $a_v=a_u\equiv a$. In this gauge the Maxwell equations for static configurations
imply the equation,
\begin{equation}
\psi^*\frac{d\psi}{dr_*}-\psi\frac{d\psi^*}{dr_*} = 0 , 
\end{equation}
which implies, in turn, that the phase of $\psi$ is constant and we can take the scalar field to 
be real valued. The remaining non-trivial Maxwell equation is
\begin{equation}
\frac{d}{dr_*}\left(e^{-2\rho-2\phi} \frac{da}{dr_*}\right)+q^2L^2\psi^2 e^{-2\phi} a=0 ,
\label{maxwell}
\end{equation}
which agrees with (\ref{sourcefree}) if $q=0$, {\it i.e.} when $\psi$ is a neutral scalar field 
that does not act as a source for the Maxwell field. 

The next step is to write equations for static configurations using the $s$ variable, in which
the metric is non-degenerate at the horizon. Since $a$ is a component of a one-form
potential, it transforms under a change of coordinates,
\begin{equation}
a(r_*)=\frac{ds}{dr_*} a(s) =2\kappa\, s\, a(s) .
\end{equation}
It follows that the potential vanishes at the horizon in tortoise 
coordinates as long as it remains a smooth function there in the $s$ variable.
The Maxwell equation becomes
\begin{equation}
\frac{d}{ds}\left(e^{-2\rho-2\phi}\left(a+ s\frac{da}{ds}\right)\right)
+\frac{q^2L^2}{4}\psi^2 e^{-2\phi} a=0 ,
\label{maxwellins}
\end{equation}
and for a smooth horizon at $s=0$ the initial data must satisfy
\begin{equation}
-a'(0)+a(0)(\rho'(0)+\phi'(0))=\frac{q^2L^2}{8}\psi(0)^2 a(0)e^{2\rho(0)} .
\end{equation}
The value of $a(0)$ is a free parameter that determines the black hole charge. 
To see this, we note that for $q=0$ we have $a+s a'=\frac{Q}{4}e^{2\phi+2\rho}$, with $Q$ the 
black hole charge, and in this case 
\begin{equation}
a(0)=\frac{Q}{4}e^{2\phi(0)+2\rho(0)}.
\label{azero}
\end{equation} 
If there is a non-vanishing scalar field with $q\neq 0$, then the solution of the Maxwell 
equation will no longer be a simple Coulomb field but we are nevertheless free to write $a(0)$
the same way as in (\ref{azero}) and the constant $Q$ can still be interpreted as the total 
charge inside the black hole.

The scalar field equation in the $s$ variable is
\begin{equation}
0=s \psi'' +\psi'-2s\phi'\psi'+4sq^2L^2a^2\psi-\frac{m^2L^2}{4}e^{2\rho}\psi ,
\label{psieqins}
\end{equation} 
with the following condition on initial data at a smooth horizon,
\begin{equation}
\psi'(0)=\frac{m^2L^2}{4}e^{2\rho(0)}\psi(0) .
\label{initialpsi}
\end{equation} 
The equation for $\tilde{f}$ remains unchanged but the remaining two field equations
get contributions from the scalar action,
\begin{eqnarray}
0&=&s\phi''+\phi'-2s\phi'^2+\frac{k}{4}e^{2\rho+2\phi}+\frac{(z^2{+}z{+}4)}{8}e^{2\rho}
-\frac{1}{4}\tilde{f}^2e^{2\rho+4\phi} \nonumber \\
&\ &\quad -4(a+s\,a')^2e^{-2\rho}-\frac{m^2L^2}{16}e^{2\rho}\psi^2, 
\label{psiphieqins}\\
0&=&s\rho''+\rho'-s\phi'^2+\frac{k}{4}e^{2\rho+2\phi}
-\frac{1}{2}\tilde{f}^2e^{2\rho+4\phi}-\frac{s}{2z}e^{4\phi}\tilde{f}'^2\nonumber \\
&\ &\quad -8(a+s\,a')^2e^{-2\rho}+\frac{s}{4}\psi'^2-sq^2L^2a^2\psi^2 ,
\label{psirhoeqins}
\end{eqnarray}
and the conditions on the initial data become,
\begin{eqnarray}
\phi'(0)&=&\frac{1}{4}\left(-ke^{2\phi(0)}{-}\frac{(z^2{+}z{+}4)}{2}
{+}(\tilde{f}(0)^2+Q^2)e^{4\phi(0)}{+}\frac{m^2L^2}{4}\psi(0)^2\right)e^{2\rho(0)}, \\
\rho'(0)&=&\frac{1}{2}\left(-\frac{k}{2}e^{2\phi(0)}
+(\tilde{f}(0)^2+Q^2)e^{4\phi(0)}\right)e^{2\rho(0)}.
\end{eqnarray}
For given values of $z$, $k$, $q$, and $m^2$ there is now a three-parameter 
family of smooth initial values for the dynamical fields at the horizon given by
$Q$, $\psi(0)$, and $f(0)$. As before, the initial data must be fine-tuned to 
obtain an asymptotically Lifshitz geometry and we take the black hole charge
$Q$ and the value of the scalar field at the horizon $\psi(0)$ as independent
parameters.

\FIGURE{\epsfig{file=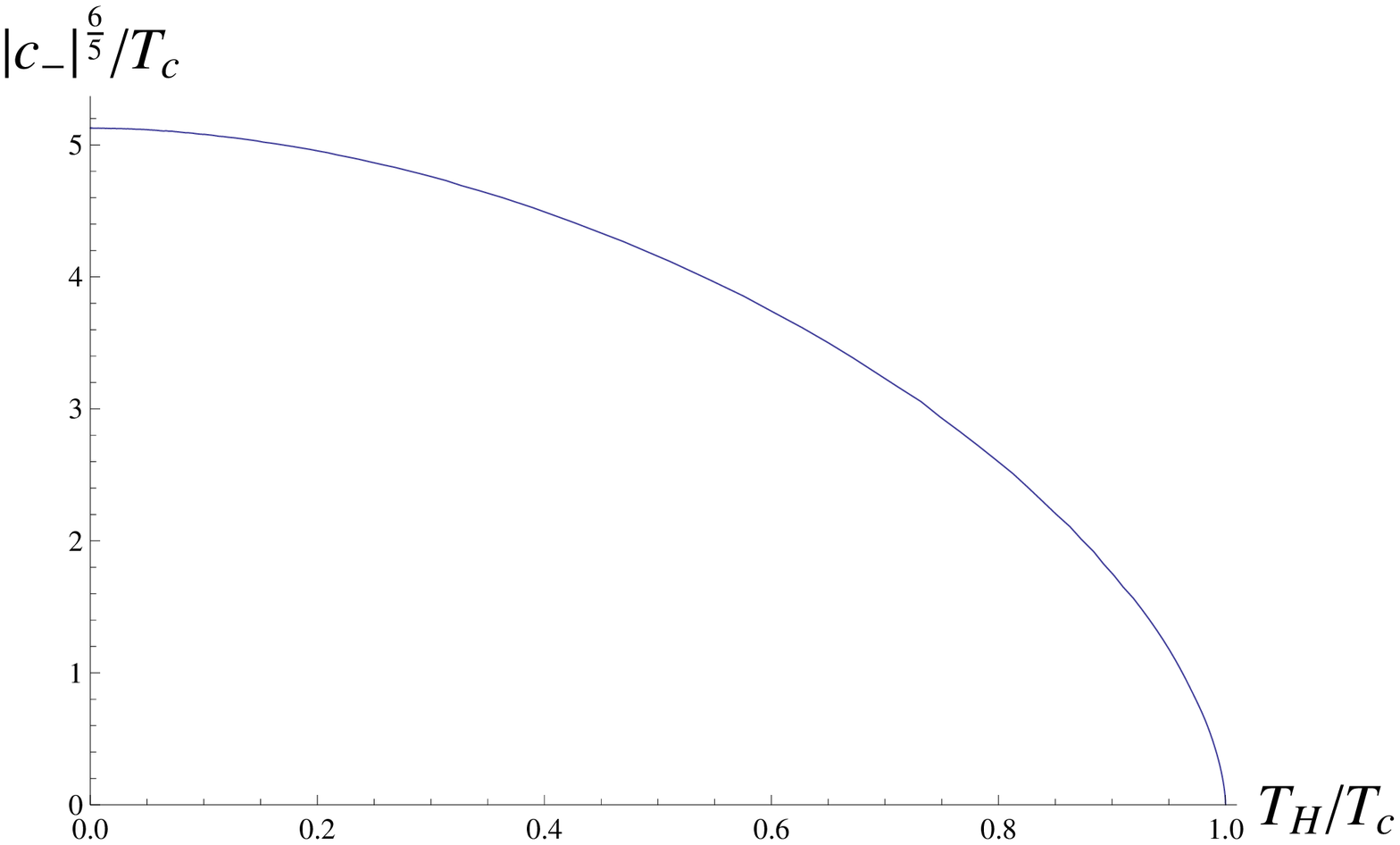,width=9cm}
\caption{Holographic superconductivity at $z=3/2$. The graph shows the 
condensate in the ${\mathcal O}_-$ theory as a function of temperature normalized to the 
critical temperature.}
\label{fig:condensates}}

\subsection{Superconducting phase}
In order to study black hole geometries that are dual to a holographic superconductor
we set $k=0$. For a given value of $z$ we set the scalar mass squared to the value in 
(\ref{mass2}) and select a value for the scalar field charge $q$. We then find numerical
black hole solutions with scalar hair for a range of $Q$ and $\psi(0)$ and study the 
asymptotic large $r$ behavior of the scalar hair in each case. The signal of a 
superconducting condensate in one or the other boundary theory is to have either 
\begin{equation}
\begin{array}{lcl}
c_+=0 & \quad\mathrm{and}\quad & \langle \mathcal{O}_-\rangle =c_- \neq 0,
\end{array}
\label{c+zero}
\end{equation}
or
\begin{equation}
\begin{array}{lcl}
c_-=0 & \quad\mathrm{and}\quad & \langle \mathcal{O}_+\rangle = c_+\neq 0.
\end{array}
\label{c-zero}
\end{equation}
The curve of vanishing $c_+$ ($c_-$) in the $Q$ {\it vs.} $\psi(0)$
plane can be tracked, tabulating the value of $c_-$ ($c_+$) as a function of the black hole
temperature along the way. Figure~\ref{fig:condensates} shows the result of this procedure
for hairy Lifshitz black holes at $z=3/2$ for scalar field charge $q=1$. There is clear evidence 
of condensation in the corresponding boundary theory. Analogous
results are obtained for other values of $q$ and more general $z>1$ but we defer a more 
detailed study of the parameter space to \cite{inprep}.

\section{Conclusions}
\label{conclusions}
In this paper we have given a holographic description of superconductors
with Lifshitz scaling. The key ingredient is a family of black holes with
scalar hair in an asymptotically Lifshitz space time.   These black holes
are interesting in their own right, and we have studied their geometry in
detail uncovering several intriguing properties, including the presence of
null singularities. Our work is an extension of previous work on Lifshitz
black holes in \cite{Danielsson:2009gi,Mann:2009yx,Bertoldi:2009vn}.

As in the case of the more conventional holographic superconductors at $z=1$,
see \cite{Gubser:2008px, Hartnoll:2008vx}, our black holes need to be equipped
with additional charges in order for a phase transition to a superconducting
phase to occur. We provide such a construction, and show how a charged
scalar in this background condenses at a critical temperature, indicating a
superconducting phase.

Our results demonstrate that phase transitions leading to superconductivity
at low temperature, can be given a holographic description also for Lifshitz
points. Such superconductors are known from solid state physics, see, e.g.
\cite{Buzdin:1984}, and we hope that our results will be useful for further 
studies of these systems. Some of their properties are different from usual
superconductors, and it would be interesting to see whether this has a
holographic counterpart.

\section{Acknowledgments}

This work was supported in part by the G\"{o}ran Gustafsson foundation, the 
Swedish Research Council (VR), the Icelandic Research Fund, the 
University of Iceland Research Fund, and the Eimskip Research Fund at the 
University of Iceland.


\bigskip

\begin{thebibliography}{99}

\bibitem{Gubser:2008px}S.~S.~Gubser,
``Breaking an Abelian gauge symmetry near a black hole horizon,''
Phys.\ Rev.\  D {\bf 78}, 065034 (2008) [arXiv:0801.2977 [hep-th]].

\bibitem{Hartnoll:2008vx}S.~A.~Hartnoll, C.~P.~Herzog and G.~T.~Horowitz,
``Building a Holographic Superconductor,'' Phys.\ Rev.\ Lett.\  {\bf 101}, 031601 
(2008) [arXiv:0803.3295 [hep-th]]
  
\bibitem {Maldacena:1997re}J.~M.~Maldacena, ``The large N limit of
superconformal field theories and supergravity,''
Adv.\ Theor.\ Math.\ Phys.\ \textbf{2}, 231 (1998)
[Int.\ J.\ Theor.\ Phys.\ \textbf{38}, 1113 (1999)] [arXiv:hep-th/9711200].

\bibitem {Hartnoll:2008kx}S.~A.~Hartnoll, C.~P.~Herzog and G.~T.~Horowitz,
``Holographic Superconductors,'' JHEP \textbf{0812}, 015 (2008)
[arXiv:0810.1563 [hep-th]].

\bibitem{Breitenlohner:1982jf}P.~Breitenlohner and D.~Z.~Freedman,
``Stability In Gauged Extended Supergravity,'' Annals Phys.\  {\bf 144}, 249 (1982).

\bibitem{Hartnoll:2009sz}S.~A.~Hartnoll,
``Lectures on holographic methods for condensed matter physics,''
Class.\ Quant.\ Grav.\  {\bf 26}, 224002 (2009) [arXiv:0903.3246 [hep-th]].

\bibitem{Herzog:2009xv}C.~P.~Herzog,
``Lectures on Holographic Superfluidity and Superconductivity,''
J.\ Phys.\ A  {\bf 42}, 343001 (2009) [arXiv:0904.1975 [hep-th]].

\bibitem{Rokhsar-1988}
  D.S.~Rokhsar and S.A.~Kivelson,
  ``Superconductivity and the Quantum Hard-Core Dimer Gas,''
  Phys.\ Rev.\ Lett.\  {\bf 61}, 2376 (1988).
  
\bibitem{Ardonne:2003wa}
  E.~Ardonne, P.~Fendley and E.~Fradkin,
  ``Topological order and conformal quantum critical points,''
  Annals Phys.\  {\bf 310}, 493 (2004)
  [arXiv:cond-mat/0311466].  

\bibitem {Fradkin-2003}
E.~Fradkin, D.A.~ Huse, R.~Moessner, V.~Oganesyan, and S.L.~Sondhi,
``Bipartite Rokhsar-Kivelson points and Cantor deconfinement,'' 
Phys.\ Rev.\ B \textbf{69} (2004) 224415 [arXiv:cond-mat/0311353].
  
\bibitem {vishwanath-2003}
A.~Vishwanath, L.~ Balents, and T.~Senthil,
``Quantum Criticality and Deconfinement in Phase Transitions Between Valence
Bond Solids,'' Phys.\ Rev.\ B \textbf{69} (2004) 224416 [arXiv:cond-mat/0311085].  

\bibitem {Kachru:2008yh}S.~Kachru, X.~Liu, and M.~Mulligan, ``Gravity Duals of
Lifshitz-like Fixed Points," Phys.\ Rev.\ D \textbf{78} (2008) 106005
[arXiv:0808.1725 [hep-th]].

\bibitem {Danielsson:2009gi}U.~H.~Danielsson and L.~Thorlacius, ``Black holes
in asymptotically Lifshitz spacetime,'' JHEP \textbf{0903}, 070 (2009)
[arXiv:0812.5088 [hep-th]].

\bibitem {Mann:2009yx}R.~B.~Mann, ``Lifshitz Topological Black Holes,''
JHEP {\bf 0906}, 075 (2009) [arXiv:0905.1136 [hep-th]].
  
\bibitem{Bertoldi:2009vn}G.~Bertoldi, B.~A.~Burrington and A.~Peet,
``Black Holes in asymptotically Lifshitz spacetimes with arbitrary critical exponent,''
Phys.\ Rev.\  D {\bf 80}, 126003 (2009) [arXiv:0905.3183 [hep-th]].

\bibitem{Bertoldi:2009dt}G.~Bertoldi, B.~A.~Burrington and A.~W.~Peet,
``Thermodynamics of black branes in asymptotically Lifshitz spacetimes,''
Phys.\ Rev.\  D {\bf 80}, 126004 (2009) [arXiv:0907.4755 [hep-th]].

\bibitem{Taylor:2008tg}M.~Taylor, ``Non-relativistic holography,'' 
arXiv:0812.0530 [hep-th].

\bibitem{Azeyanagi:2009pr}T.~Azeyanagi, W.~Li and T.~Takayanagi,
``On String Theory Duals of Lifshitz-like Fixed Points,'' JHEP {\bf 0906}, 084 (2009)
[arXiv:0905.0688 [hep-th]].

\bibitem{Pang:2009ad} D.~W.~Pang, ``A Note on Black Holes in Asymptotically 
Lifshitz Spacetime,'' arXiv:0905.2678 [hep-th].

\bibitem{Birnir:1992by}B.~Birnir, S.~B.~Giddings, J.~A.~Harvey and A.~Strominger,
``Quantum Black Holes,'' Phys.\ Rev.\  D {\bf 46}, 638 (1992) [arXiv:hep-th/9203042].

\bibitem{Ross:2009ar}S.~F.~Ross and O.~Saremi, ``Holographic stress tensor for 
non-relativistic theories,'' JHEP {\bf 0909}, 009 (2009) [arXiv:0907.1846 [hep-th]].

\bibitem{Klebanov:1999tb}I.~R.~Klebanov and E.~Witten,
``AdS/CFT correspondence and symmetry breaking,''
Nucl.\ Phys.\  B {\bf 556}, 89 (1999) [arXiv:hep-th/9905104].

\bibitem{inprep} E.J.~Brynjolfsson, U.H.~Danielsson, L.~Thorlacius, and T.~Zingg,
in preparation.

\bibitem{Buzdin:1984} A.I.~Buzdin and M.L.~Kulic, 
``Unusual behaviour of superconductors near the tricritical Lifshitz point,'' 
Journal of Low Temperature Physics, {\bf 54} 203 (1984).


\end{thebibliography}
\end{document}